\documentclass[aps,prl,reprint,superscriptaddress,showpacs,amsmath,amssymb,longbibliography,lengthcheck]{revtex4-1}
\usepackage{graphicx}
\usepackage{mathrsfs}
\usepackage{txfonts}

\begin{document}

 \title{Linear chain structure of four-$\alpha$ clusters in $^{16}$O} 

\author{T. Ichikawa}%
\affiliation{Yukawa Institute for Theoretical Physics, Kyoto University,
Kyoto 606-8502, Japan}
\author{J. A. Maruhn}
\affiliation{Institut fuer Theoretische Physik, 
Universitaet Frankfurt, D-60438 Frankfurt, Germany}
\author{N. Itagaki}
\affiliation{Yukawa Institute for Theoretical Physics, Kyoto University,
Kyoto 606-8502, Japan}
\author{S. Ohkubo}
\affiliation{Department of Applied Science and Environment, 
University of Kochi, Kochi 780-8515, Japan}
\affiliation{Research Center for Nuclear Physics, Osaka University, Ibaraki, Osaka 567-0047, Japan}
\date{\today}

\begin{abstract}
We investigate the linear-chain configurations of four-$\alpha$ clusters
in $^{16}$O using a Skyrme cranked Hartree-Fock method and discuss the
relationship between the stability of such states and angular momentum.
We show the existence of a region of angular momentum (13-18 $\hbar$)
where the linear chain configuration is stabilized.  For the first time
we demonstrate that stable exotic states with a large moment of inertia
($\hbar^2/2\Theta$ $\sim$ 0.06-0.08 MeV) can exist.
\end{abstract}


\maketitle

Strong nuclear deformations provide an excellent framework in which to
investigate the fundamental properties of quantum many-body
systems. Strongly deformed nuclei have been identified by the
observation of $\gamma$-ray cascades typical of rotational bands.  Since
the first observation of such bands \cite{Nyako}, strongly deformed
states with an aspect ratio 1:2 have been found in various nuclei.
These bands are called superdeformed bands.  Furthermore, the
hyperdeformed bands, in which the deformation is around 1:3, have been
reported in several experiments \cite{Galindo-Uribarri}.  At first,
those strongly deformed states were found in the heavy nuclei. Such new
data have triggered interest in whether more exotic states exist in
light nuclei where a strong deformation above 1:3 could be possible due
to $\alpha$-cluster structure.

 Experimental candidates for strongly deformed states with an aspect
ratio above 1:3 have been suggested in light $4N$ nuclei.  One candidate
is the four-$\alpha$ linear chain band starting around the four-$\alpha
$ threshold energy region in $^{16}$O suggested by Chevallier {\em et\
al.}~\cite{Chevallier} in the $^{12}$C ($\alpha$, $^8$Be)$^8$Be reaction
and was supported by Suzuki {\em et\ al.}~\cite{Suzuki}.  Freer {\em et\
al.}~\cite{Freer1995} performed the $^{12}$C($^{16}$O,4$\alpha$)
reaction and obtained a smaller moment of inertia, about $2/3$ of
Ref.~\cite{Chevallier}.  Recently it has been suggested that these have
a loosely coupled four-$\alpha $ structure~\cite{Ohkubo2010}, in
connection with the gas-like $0^+_2$ (Hoyle) state in $^{12}$C
\cite{Uegaki1977}.  Another candidate is a six-$\alpha$ linear chain
state, which has been extensively studied both theoretically and
experimentally \cite{Merchant1992,Rea,Zamick1992,Wuosmaa}.  Wuosmaa {\em
et\ al.}~\cite{Wuosmaa} and Rae {\em et\ al.}~\cite{Rea} suggested that
the molecular resonance state observed in the inelastic reaction
$^{12}$C($^{12}$C, $^{12}$C($0^+_2$))$^{12}$C($0^+_2$) might be a
candidate for the six-$\alpha$ linear chain state.  Hirabayashi {\em et\
al.}~claimed \cite{Hirabayashi} that this has a loosely coupled
3$\alpha$+3$\alpha$ configuration rather than a linear chain.  The
seven-$\alpha$ linear chain state in $^{28}$Si was not
observed~\cite{Simmons1995}.  Despite many efforts no clear experimental
evidence of a stable $\alpha$ linear chain structure has been confirmed
and its existence remains an open problem.

The stability of such linear chain states has often been studied through
the analysis of small vibrations around the equilibrium configuration
and with the axial symmetry~\cite{Horiuchi,Zamick1992}.  However, it was
shown that bending motion is an essential path for the transition to
low-lying states \cite{Umar}.  Thus it is necessary to calculate the
stability in a wide model space, which contains lower excited states.
Two mechanisms are important for stabilizing the linear chain state.
The first mechanism is the quantum-mechanical orthogonality condition to
other low-lying states.  The second is, as discussed by Wilkinson
\cite{Wilkinson1986}, the competition between the nuclear attractive and
centrifugal forces due to rotation of the system: a large moment of
inertia such as in the linear-chain configuration is favored with a
large angular momentum.  On the other hand, high angular momenta would
lead to fission of the linear chain due to the strong centrifugal force.
Detailed investigations are necessary for the existence of a region of
angular momentum where the linear chain configuration is stabilized.

In this Letter we show that a region of angular momentum (13-18 $\hbar$)
where the four-$\alpha$ linear chain configuration is stabilized exists
in $^{16}$O.  The cranked Hartree-Fock (HF) method is used to
investigate the stability of the configuration and the moment of
inertia.  The mechanism of the stabilization against the decay with
respect to bending motion and fission and its angular-momentum
dependence is clarified.

Until now, most of the theoretical analyses of the linear chain
structure have been performed using the conventional cluster model with
effective interactions, whose parameters are determined to reproduce the
binding energies and scattering phase shifts of the clusters. Thus it is
highly desirable to study the presence of exotic cluster configurations
based on different approaches, such as mean field models.  The effective
interactions used in the mean field models are determined in a
completely different way; they are designed to reproduce various
properties of nuclei in a wide mass range.  The appearance of cluster
structure as a result of calculations with such interactions and model
spaces would give more confidence in their presence.  Recent
developments of three-dimensional calculations with Skyrme forces enable
us to describe both shell-like and cluster-like configurations, and the
interplay between these structures based on this approach has been
successfully investigated \cite{Maruhn}.

\begin{figure}[t]
\includegraphics[keepaspectratio,scale=0.18]{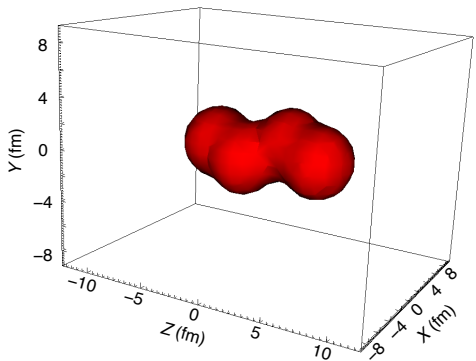}\\%
\caption{\label{init} (Color online) Surface of the total nucleon
density distribution for the initial twisted wave function. We chose the
surface as the half of the total density.}
\end{figure}

\begin{figure*}[t]
\includegraphics[keepaspectratio,scale=0.6]{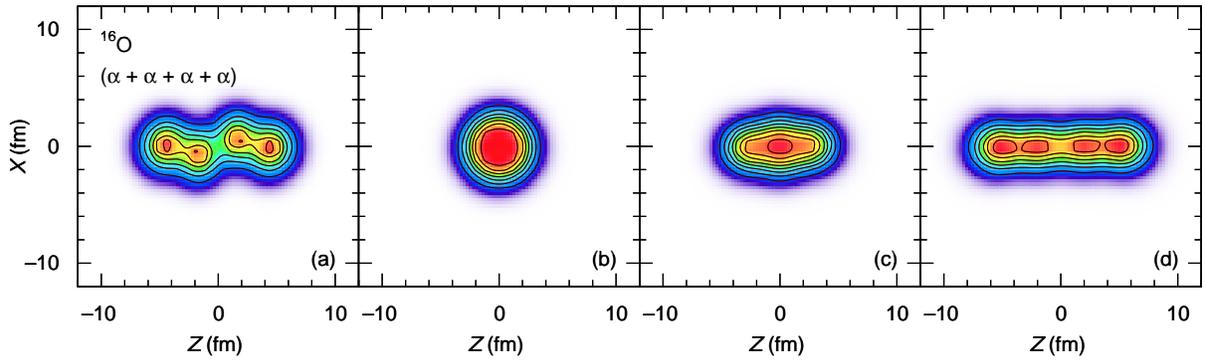}\\%
\caption{\label{den1} (Color online) Total nucleon density distribution
calculated using the cranking method for (a) the initial wave function,
(b) the ground state, (c) the quasi-stable state, and (d) the
four-$\alpha$ linear chain state.  The isolines correspond to multiples
of 0.02 fm$^{-3}$. We normalize the color to the density distribution at
the maximum of each plot.}
\end{figure*}

To investigate the four-$\alpha$ linear chain state in the rotational
frame, we perform cranked HF calculations. We self-consistently
calculate the cranked HF equation, given by $\delta\left<H-\omega
J\right>=0$, where $H$ is the total Hamiltonian, $\omega$ is the
rotational frequency, and $J$ is the angular momentum around the $y$
axis.  We represent the single-particle wave functions on a Cartesian
grid with a grid spacing of 0.8~fm. The grid size is typically $24^3$
for ground states and $32\times24^2$ for superdeformed states. This
accuracy was seen to be sufficient to provide converged configurations.
The numerical procedure is the damped-gradient iteration method
\cite{gradient}, and all derivatives are calculated using the Fourier
transform method.

We take three different Skyrme forces which all perform very well
concerning nuclear bulk properties but differ in details: Sly6 as a
recent fit which includes information on isotopic trends and neutron
matter \cite{Cha97a}, and SkI3 as well as SkI4 as recent fits which map
the relativistic isovector structure of the spin-orbit force
\cite{Rei95a}.  The force SkI3 contains a fixed isovector part analogous
to the relativistic mean-field model, whereas SkI4 is adjusted allowing
free variation of the isovector spin-orbit term. Thus all forces differ
somewhat in their actual shell structure. Besides the effective mass,
the bulk parameters (equilibrium energy and density, incompressibility,
and symmetry energy) are comparable.

Here we discuss the stability of the four-$\alpha$ linear chain
configuration in the rotating frame for $^{16}$O.  To this end, we
perform the cranked HF calculations with various rotational frequencies,
$\omega$. For the initial wave function, we chose the z-axis as the
principal axis and use the twisted four-$\alpha$ configuration, as shown
in Fig.~\ref{init}. We also show the corresponding two-dimensional plot
in Fig.~\ref{den1} (a).  Note that this initial is a three-dimensional
four $\alpha$ configuration, which facilitates the transition of the
initial state to low-lying states including the ground state during the
convergence process. This was demonstrated for the carbon chain states
in Refs.~\cite{Itagaki-C,Maruhn-C}.  We calculate the rigid-body moment
of inertia, $\Theta$, using the total nucleon density at each iteration
step.  We only consider rotation around the $y$ axis (perpendicular to
the deformation axis $z$).

\begin{figure}[b]
\includegraphics[keepaspectratio,width=\linewidth]{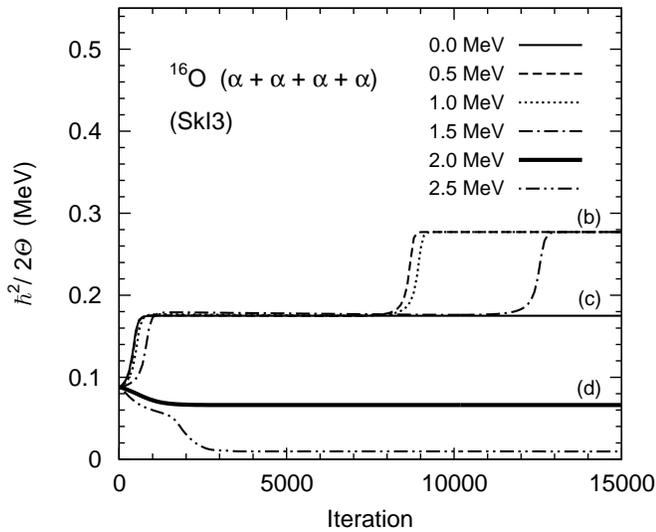}\\%
\caption{\label{ita1} Coefficient of the rotational energy,
$\hbar^2/2\Theta$, calculated using the cranking method versus the HF
iterations with various rotational frequencies $\omega$. The symbols
(b), (c), and (d) correspond to the density distributions given in
Fig.~\ref{den1}.}
\end{figure}

We first investigate the convergence behavior of the HF iterations.  To
check this, we calculate the coefficient of the rotational energy, given
by $\hbar^2/2\Theta$, at each iteration step.  Figure~\ref{ita1} shows
the calculated results with various rotational frequencies versus the
iterations in the case of the SkI3 interaction.  The initial state with
the twisted linear chain configuration is not the true ground state of
the HF model space and the solution changes into the true ground state
after some large number of iterations; however the situation depends on
the value of the rotational frequency $\omega$.  In Fig.~\ref{ita1}, we
see that the rotational frequencies $\omega=0.5$, 1.0, and 1.5
MeV/$\hbar$ (the dashed, dotted, and dot-dashed lines, respectively)
lead to the true ground state. Note that the rigid-body moment of
inertia, $\Theta$, is a classical value, which does not become to be
zero even with spherical shapes. The corresponding density distribution
at the 15000th iteration is plotted in Fig.~\ref{den1} (b).  The
frequency $\omega=0.0$ MeV/$\hbar$ (the solid line in Fig.~\ref{ita1})
leads to the quasi-stable state (see Fig.~\ref{den1} (c)).  At around
$\omega=2.0$ MeV/$\hbar$, we obtain the state (the thick solid line in
Fig.~\ref{ita1}) with the four-$\alpha$ linear chain configuration, as
shown in Fig.~\ref{den1} (d), whereas fission occurs above those
rotational frequencies (the dot-dot-dashed line in Fig.~\ref{ita1}).

\begin{figure}[htbp]
\includegraphics[keepaspectratio,width=\linewidth]{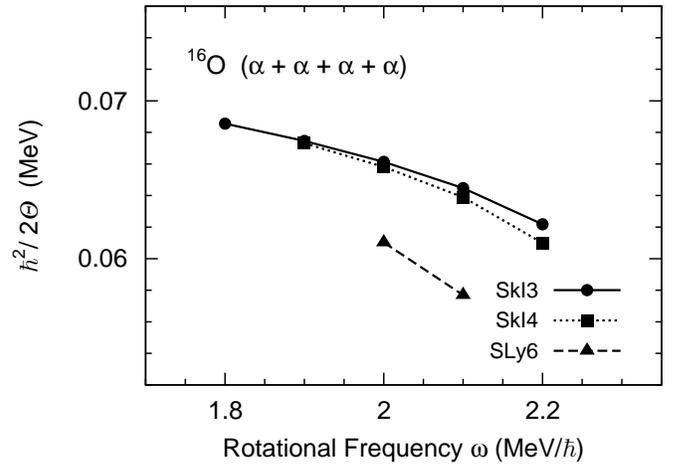}\\%
\caption{\label{iner} Coefficient of the rotational energy
$\hbar^2/2\Theta$ as a function of rotational frequency $\omega$.  The
lines correspond to the different Skyrme forces as indicated.}
\end{figure}

We next estimate the range of the rotational frequencies where the
four-$\alpha$ linear-chain configuration can be stabilized.
Figure~\ref{iner} shows the coefficient of the rotational energy,
$\hbar^2/2\Theta$, versus the rotational frequency $\omega$ with various
Skyrme interactions.  We find stable states for the four-$\alpha$ linear
chain configuration for all of the interactions.  For the SkI3
interaction, we obtain the lower and upper bounds of the rotational
frequencies as 1.8 and 2.2 MeV/$\hbar$. Between these the four-$\alpha$
linear chain configuration is stabilized.  The values are 1.9 and 2.2
MeV/$\hbar$ for the SkI4 interaction and 2.0 and 2.1 MeV/$\hbar$ for the
SLy6 interaction, respectively.  In these frequency regions where the
linear chain configuration is stabilized, we can define the rigid-body
moments of inertia, which are calculated as 0.065 MeV for the SkI3 and
SkI4 interactions and 0.06 MeV for the SLy6 interaction.  These values
are consistent with 0.063 in a naive picture of rigid-body
four-$\alpha$'s laid in linear chain with 12 fm as in Fig.~2(d).

\begin{figure}[htbp]
\includegraphics[keepaspectratio,width=\linewidth]{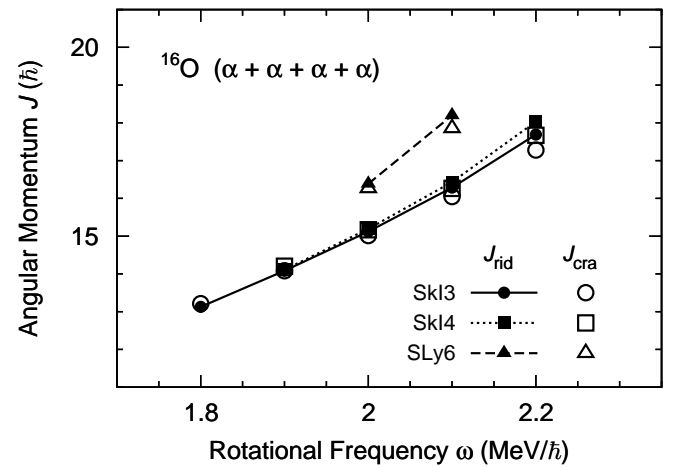}\\%
\caption{\label{ang} Angular momentum as a function of rotational
frequency $\omega$ for the Skyrme forces.  The lines with solid symbols
denote the calculated results for the rigid-body moment of inertia,
while the open symbols denote the results for the cranking method.  }
\end{figure}

We also estimate the corresponding angular momentum where the
four-$\alpha$ linear chain configuration is stabilized. We calculate the
angular momentum using the rigid-body moment of inertia obtained and
compare it with the value calculated by the cranking method. The angular
momentum with the rigid-body moment of inertia, $J_{\rm rid}$, is
calculated as $J_{\rm rid}=\Theta \omega$.  The angular moment
calculated using the cranking method, $J_{\rm cra}$, is given by $J_{\rm
cra}=<J>$, where $<J>$ is the expectation value of the angular momentum
in the cranking equation.  Figure~\ref{ang} shows the angular momentum
obtained versus the rotational frequency.  We see that the calculated
angular momentum using the rigid-body moment of inertia agrees well with
that of the cranking method, indicating that the rigid-body
approximation is reasonable for the four-$\alpha$ linear chain states.
We find that the lower and upper bounds of the angular momentum where
the four-$\alpha$ linear chain configuration is stabilized are about 13
and 18 $\hbar$ for the SkI3 interaction, 14 and 18 $\hbar$ for the SkI4
interaction, and 16 and 18 $\hbar$ for the SLy6 interaction,
respectively.  With a such high angular momentum, very exotic
configuration of the four-$\alpha$ linear chain can be stabilized.
Fission occurs beyond this angular momentum region.  Furthermore, it is
possible that states with even lower angular momenta are stabilized,
when the coupling effect with low-lying states is taken into account.

\begin{figure}[htbp]
\includegraphics[keepaspectratio,width=\linewidth]{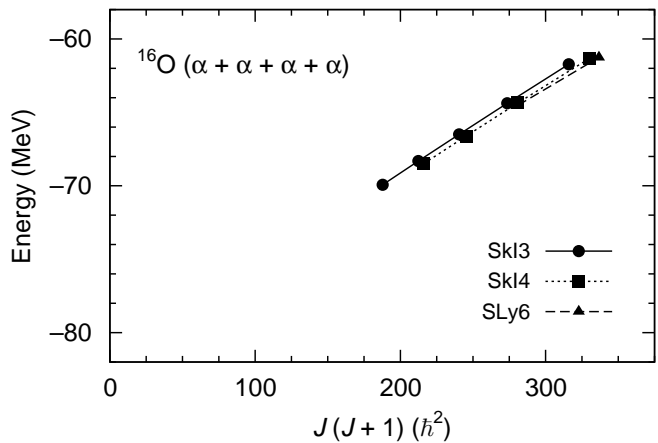}\\%
\caption{\label{ex-j} Calculated energies of the four $\alpha$
linear-chain states versus the angular momentum.  The lines correspond
to the different Skyrme forces as indicated.  }
\end{figure}

Finally in Fig.~\ref{ex-j} we show the energies of calculated four
$\alpha$ linear chain states versus the angular momentum $J$.  The
energy is very high, at around $-70$ MeV with $J \sim 14$.  If we
extrapolate the energy to lower angular momentum, the band head energy
at $J = 0$ is estimated to be around $-82$ MeV, which is much higher
than the value suggested in Ref. \cite{Chevallier}.  This rather high
excitation energy in the Skyrme forces is also reported in
Ref.~\cite{Zamick1992}.

In summary, we have investigated the stability of the four-$\alpha$
linear chain configuration in $^{16}$O using the Skyrme mean field
method with cranking. Even if the bending path is opened in the
three-dimensional space, we obtained regions of rotational frequency
where the linear chain configuration is stabilized.  Below this region,
the state converges to low-lying configurations, with fission occurring
beyond this region.  The frequency range corresponds to angular momenta
of 13-18~$\hbar$.  Furthermore, when the coupling effect with low-lying
states is taken into account there is a possibility that states with
even lower angular momentum are stabilized.  The coefficient of the
rotation ($\hbar^2/2\Theta$) of the four-$\alpha$ linear chain
configuration obtained is around 0.06-0.08 MeV.  We have, for first
time, shown that states with such large moments of inertia are possible
in light nuclei under conditions of large angular momenta.

As shown in this Letter, the exotic four-$\alpha$ linear-chain state can
indeed exist in $^{16}$O.  We also investigated whether such state can
be accessible in the $^{8}$Be + $^{8}$Be fusion reaction using the
time-dependent HF method with the same Skyrme force as the present
study~\cite{fusion11}. We obtained a quasi-stable state with a similar
moment of inertia, as shown here.  The HF method is a powerful tool for
investigating both the static and dynamical properties of nuclei in the
consistent framework.  This method is a promising tool to reveal the
existence of states with more exotic geometric configurations, such as
longer linear chain~\cite{6a} and polygons~\cite{Wilkinson1986}, which
is still an open question for their existence.

\begin{acknowledgments}
  This work was undertaken as part by the Yukawa International Project for
  Quark-Hadron Sciences (YIPQS), and was partly supported by the GCOE
  program ``The Next Generation of Physics, Spun from Universality and
  Emergence'' from MEXT of Japan.  J.A.M. was supported by the
  Frankfurt Center for Scientific Computing and by the BMBF under
  contract 06FY9086.  One of the authors (JAM) would like to thank the
  Japan Society for the Promotion of Science (JSPS) for an invitation
  fellowship for research in Japan.
\end{acknowledgments}

\end{document}